# How Turbulent Jets Can Disperse Virus Clouds in Poorly Ventilated Spaces

J. Tahmassebpur[1] and P. H. Diamond[2]

[1] Department of Electrical and Computer Engineering, University of California, San Diego, La Jolla, United States
[2] Department of Physics, University of California, San Diego, La Jolla, United States

E-mail: jt577@cornell.edu, pdiamond@ucsd.edu

**Abstract**
We show that enhanced turbulent mixing can be used to mitigate airborne COVID-19 transmission by dispersing virus-laden clouds in enclosed, poorly ventilated spaces. A simple system of fan-driven turbulent jets is designed so as to minimize peak concentrations of passive contaminants on time scales short compared to the room ventilation time. Standard Reynolds-average and similarity methods are used, and combinations of circular and radial wall jets are considered. The turbulent diffusivity and contaminant mixing time are calculated. Results indicate that this approach can significantly reduce peak virus cloud concentrations, especially in small spaces with low occupancy, such as restrooms. Turbulent mixing is, of course, ineffective in the absence of ventilation.

## 1. Introduction

The Coronavirus disease (COVID-19), identified as the severe acute respiratory syndrome SARS-CoV-2 (Gorbalenya *et al* 2020), has spread across the world infecting millions over the past two years. Major venues for the spread of this disease are confined indoor spaces like bathrooms, hospitals, offices, etc. Hence, it is crucial to understand how the virus is transmitted in these spaces. The three main modes of transmission are: direct contact with bodily fluids, the spraying of large droplets directly onto body fluids, and the inhalation of small, airborne aerosols (Bourouiba *et al* 2014). Thus far, society has mainly taken precaution with mostly the first two modes (social distancing, sanitizing, masks, etc.), but there is evidence that airborne transmission is even more dangerous. Studies showed that superspreading events like choir practice, restaurants, gatherings, etc. are dangerous because of the airborne transmission of aerosols in conjunction with poor ventilation (Morawska *et al* 2020, Lu *et al* 2020). This is because violent expirations like coughing and sneezing release hot, buoyant, turbulent air that suspends droplets of various sizes and vectors them across the room. Larger droplets with diameter $d > 100\mu m$ settle to the ground within a second, following ballistic trajectories (Wells 1934), while droplets with $d < 100\mu m$ have settling speeds $< 3mm/s$ (Wells 1955), suspending them in the air for 1-2 hours. These suspended droplets are important for long range transmission. Studies also show that given the same viral load, aerosols are more infectious than larger droplets because they penetrate more deeply into the respiratory tract (Sonkin 1951, Wells 1955). In terms of inhalation, droplets with $d > 20\mu m$ are unlikely to reach the pulmonary region at all, so they should not be considered as a risk (Nicas *et al* 2005). Although droplets with $d < 20\mu m$ constitute over 60% of all droplets in most violent expiratory events (Duguid 1946), only 1% of them carry significant viral load (Stadnytskyi *et al* 2020) so it is likely that only high emitters contribute to airborne transmission (Anand *et al* 2020).

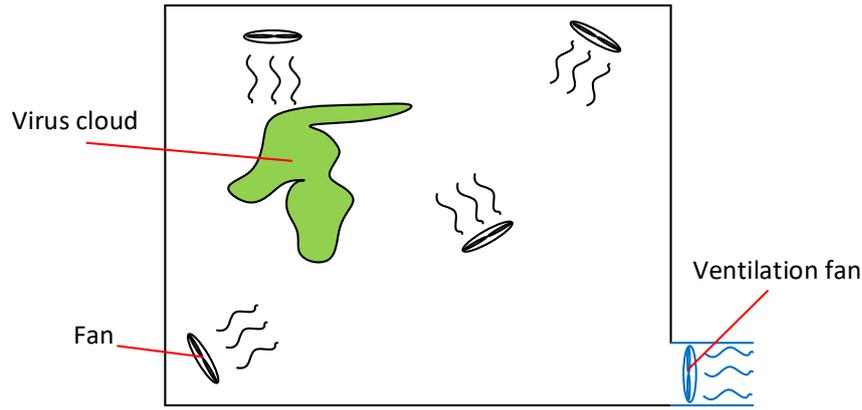

**Figure 1:** Depiction of a ventilated space with additional fans diffusing a virus cloud.

Infections in enclosed spaces should therefore most often occur due to airborne transmission by virus clouds created by high emitters. These virus clouds can remain in a room indefinitely unless proper ventilation is in place. Ventilation systems usually have ventilation times (the time it takes to completely change the air in the room) anywhere from 3-60 minutes (Riediker 2020). Within this timeframe, bystanders in the room may be infected since the virus cloud created by a high emitter will remain in a relatively small volume at high concentrations. This is more dangerous than if the cloud were fully mixed in the room, because in the former case, a bystander could encounter the highly concentrated cloud and immediately inhale a significant dose. However, the critical dose of SARS-CoV-2 is difficult to determine and could vary significantly. A study conducted on ferrets found that a dose of 500 virus particles resulted in only one of six ferrets experiencing viral RNA replication, indicating a threshold for immunity (Ryan *et al* 2020). Since ferrets have similar pathology in the lung to humans (Brand *et al* 2008), we will take this number as indicative of humans as well. Some works oppose mixing because it exposes more people to the virus as it spreads over a great distance (Mingotti *et al* 2020), however they do not consider how the concentration decreases as it spreads. The key question this paper hopes to address is: how might we trade a high concentration of virus in a small space for a sufficiently low concentration in the whole space?

| Space Type | Time of Exposure, $t_{ex}$ (hr) | Ventilation Time, $t_v$ (min) | Time Delay, $t_{del}$ (s) | Mixing Time, $t_M$ (min) |
|---|---|---|---|---|
| **Classroom** | 1 | 45 | <5 | <5 |
| **Restaurant** | 1 | 30 | <10 | <5 |
| **Auditorium** | 2 | 100 | <10 | <10 |
| **Laboratory** | 4 | 50 | <5 | <5 |
| **Plane Cabin** | 5 | 20 | <5 | <2 |

**Table 1**: Table of how time scales should look like for different types of poorly ventilated spaces

To address this, we must consider the timescales involved: the time delay, $t_{del}$, the mixing time, $t_M$, the time of exposure, $t_{ex}$, and the ventilation time, $t_v$. The time delay is the time it takes for the turbulence to homogenize in the space. In other words, it is the time elapsed before mixing begins. The mixing time is how long it takes the turbulence to mix a virus cloud of high concentration to a specified low concentration. Both $t_{del}$ and $t_M$ are controlled by the parameters defining the apparatus driving the turbulence. On the other hand, $t_{ex}$, which is the time that a bystander is expected to be in the space, and $t_v$ are predefined quantities. For turbulent mixing to be of use, the ordering of time scales should satisfy $t_{del} \ll t_M \ll t_v, t_{ex}$, so that the cloud is fully mixed before the ventilation time elapses. **Table 1** shows desirable values of $t_{del}$ and $t_M$ for given $t_{ex}$ and $t_v$ in some common spaces.

It should be noted that mixing is not a substitute for ventilation. Both must work together to lower the peak concentrations of virus clouds and remove the infected air from the space, otherwise concentrations will continue to rise over time if there are high emitters present. An example situation where mixing alone fails is the superspreading event in Hawaii where a spin class of 10 people were all infected with COVID-19 (Parker-Pope 2021). The instructor kept the windows and doors closed but placed three large fans in the room to keep people cool. Although the air was constantly being mixed, everyone in the class tested positive because there was no ventilation mechanism to remove the contaminated air.

In this paper, a model of the mixing of virus clouds in a space with an arbitrary arrangement of fans is developed. The turbulent jets emitted by fans are modelled by expressions that use familiar results from turbulence modelling and prior experiments. Two types of jets will be considered: circular jets (those emitted directly by the fan) and radial wall jets (created when a circular jet impinges on a wall of the space). Assuming the turbulence is homogenous throughout the space for $t > t_{del}$, the turbulent diffusion coefficient for cloud particles is calculated for an arbitrary fan arrangement. Using Fick's law, the mixing time of one virus cloud produced by the cough of a high emitter is then determined. The analytical model developed demonstrates that fan arrangements like that in **Figure 1** can mix virus clouds on times significantly shorter than poor (i.e. long) ventilation times, potentially reducing the risk of infection for bystanders in certain cases if they are in the space for no longer than $t_{ex}$.

The remainder of the paper is organized as follows. In part two, a model for the mean velocity and turbulent kinetic energy of jets produced by fans in the space is introduced. In part three, these models are used to derive the diffusivity for an arbitrary fan arrangement. In part four, the equation for mixing time for a virus cloud created by the cough of a high emitter is derived. In part five, applications and limitations of the model are discussed.

## 2. Model Derivation

### *2.1 Mean Velocity and Turbulent Kinetic Energy of Jets*

A fan initially produces a turbulent circular jet as depicted in **Figure 2**, which then changes form as it interacts with walls, surfaces, or other fans' jets. Before investigating subsequent forms, a model for the circular jet will be developed. By simplifying the Reynolds Averaged Navier Stokes (RANS) equations to only account for evolution in the longitudinal ($z$) direction, the mean velocity of a circular jet may be defined as

$$U_c \frac{\partial U_c}{\partial z} = -\frac{1}{r}\frac{\partial}{\partial r}(r\langle uv \rangle) \tag{1}$$

where $r$ is the radial coordinate and $\langle uv \rangle$ is the Reynolds stress. The Reynolds stress may be expressed in terms of the mixing length, $\delta = \delta(z)$, and the mean velocity by

$$\langle uv \rangle = -\delta^2 \frac{\partial \langle U_c \rangle}{\partial r}\left|\frac{\partial \langle U_c \rangle}{\partial r}\right| \tag{2}$$

Assuming the jet is self-preserving, the mean velocity takes the form

$$U_c = U_m(z)f(\xi) \tag{3}$$

where $U_m(z)$ is the velocity at the center of the jet and $f(\xi)$ is a non-dimensional function describing the shape of the mean velocity about the center. The variable $\xi = r/\delta$ denotes the self-preserving spread of the circular jet. Substituting Eqns. 2 and 3 into Eq. 1 gives

$$\left[U_m \frac{dU_m}{dz}\right]f^2 - \left[U_m^2 \frac{1}{\delta}\frac{d\delta}{dz}\right]\xi f \frac{df}{d\xi} = \left[\frac{U_m^2}{\delta}\right]\frac{1}{\xi}\left(\left(\frac{df}{d\xi}\right)^2 + 2\xi \frac{df}{d\xi}\frac{d^2 f}{d\xi^2}\right) \tag{4}$$

For the jet to be self-preserving, the bracketed terms must be of the same order in $z$. This order cannot be derived from Eq. 4 alone. The constancy of momentum flux must also be used. By integrating both sides of Eq. 1 and invoking the boundary conditions $\frac{\partial \langle U \rangle}{\partial r} = 0$ at $r = 0, \infty$ the momentum flux takes the form

$$[U_m^2 \delta^2]\int_0^\infty f^2 \xi d\xi = const \sim z^0 \tag{5}$$

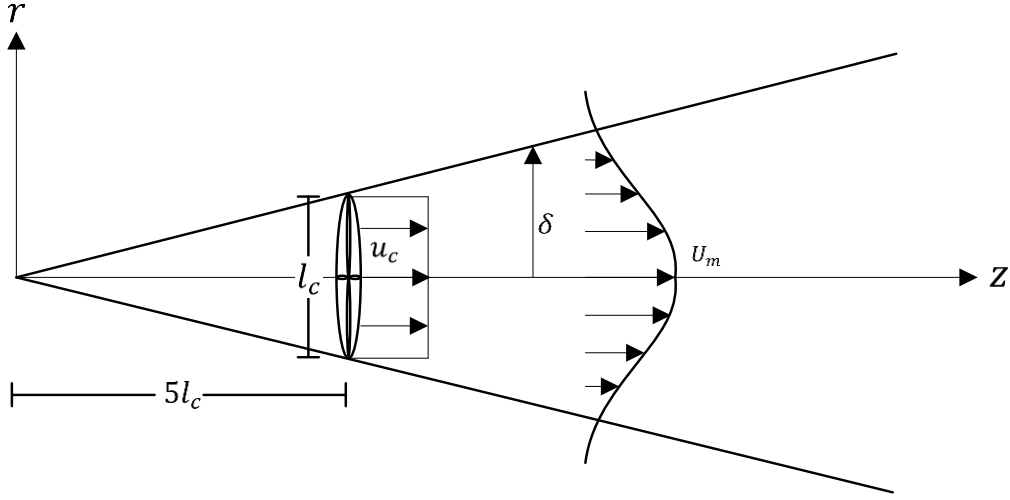

**Figure 2:** A circular jet emanating from a virtual point source. The real initial position of the jet is at $z = 5l_c$.

where $[U_m^2 \delta^2]$ must be of order $z^0$ for the momentum flux to be constant. If $U_m \sim z^n$ and $\delta \sim z^m$, then $n = -m$. Using the relationship between the bracketed terms in Eq. 4, the values of $n$ and $m$ are found to be $-1$ and $1$, respectively. This yields the following estimates for the evolution of the mixing length and central velocity: $\delta \sim z, U_m \sim 1/z$. From experiments in (Rajaratnam 1976) the shape of $f(\xi)$ can be modelled as a Gaussian, yielding the mean velocity

$$U_c = u_c \frac{5l_c}{z} e^{-70\xi^2} \tag{6}$$

where $u_c$ is the initial speed of the jet and $l_c$ is the initial size of the jet. In this equation, $\xi = r/z$. The constant in the argument of the Gaussian fits the experimentally determined mixing length $\delta = 0.1z$ (Rajaratnam 1976). Since the model assumes that the jet emanates from a point source, the jet must be normalized so that it has initial size $l_c$. Therefore, the actual jet begins at $z = 5l_c$. Note that this value does not mean the jet is created $z = 5l_c$ away from the fan; the fan would also be at $z = 5l_c$.

The turbulent kinetic energy of the circular jet, defined as $k_c = \frac{1}{2}(\langle u \rangle^2 + \langle v \rangle^2)$ in 2 dimensions, may be approximated as:

$$U_c \frac{\partial k_c}{\partial z} = -\frac{1}{r}\frac{\partial}{\partial r}(r\langle k_c v \rangle) - \langle uv \rangle \frac{\partial U_c}{\partial r} - \varepsilon \tag{7}$$

where $\varepsilon$ is the dissipation term. The dissipation has the form $\varepsilon = \frac{U_m^3}{\delta} D(\xi)$, where $D(\xi)$ describes the arbitrary self-preserving shape of the dissipation. Although an analytical form of $D(\xi)$ can be found by using a $k - \varepsilon$ model, it is unnecessary to include it in this work because we are only interested in deriving the order of magnitude behavior. The self-preserving shape of the turbulent kinetic energy can be approximated from experimental results. The quantity $\langle k_c v \rangle = -\frac{\nu_T}{\sigma_k}\frac{\partial k}{\partial r}$ with $\sigma_k = 1$ (Pope 2000). Taking $k_c = K_m(z)g(\xi)$, where $K_m(z)$ is the turbulent kinetic energy at the center of the jet, and $g(\xi)$ is a non-dimensional function describing the shape of the turbulent kinetic energy about the center, Eq. 7 becomes

$$\left[U_m \frac{dK_m}{dz}\right]fg - \left[U_m K_m \frac{1}{\delta}\frac{d\delta}{dz}\right]f\xi\frac{dg}{d\xi} = \left[\frac{U_m K_m}{\delta}\right]\frac{1}{\xi}\left(\frac{df}{d\xi}\frac{dg}{d\xi} + \frac{d^2f}{d\xi^2}\frac{dg}{d\xi}\xi + \frac{d^2g}{d\xi^2}\frac{df}{d\xi}\xi\right) + \left[\frac{U_m^3}{\delta}\right]\left(\left(\frac{df}{d\xi}\right)^3 - D(\xi)\right) \tag{8}$$

Once again, self-preservation forces the bracketed terms to be of the same order in $z$. Using $\delta \sim z, U_m \sim 1/z$, one can deduce that $K_m \sim 1/z^2$. Experimental observations in (Lai 2019) suggest that the shape of the turbulent kinetic energy may be approximated as a Gaussian with the same spread coefficient as the mean velocity, yielding

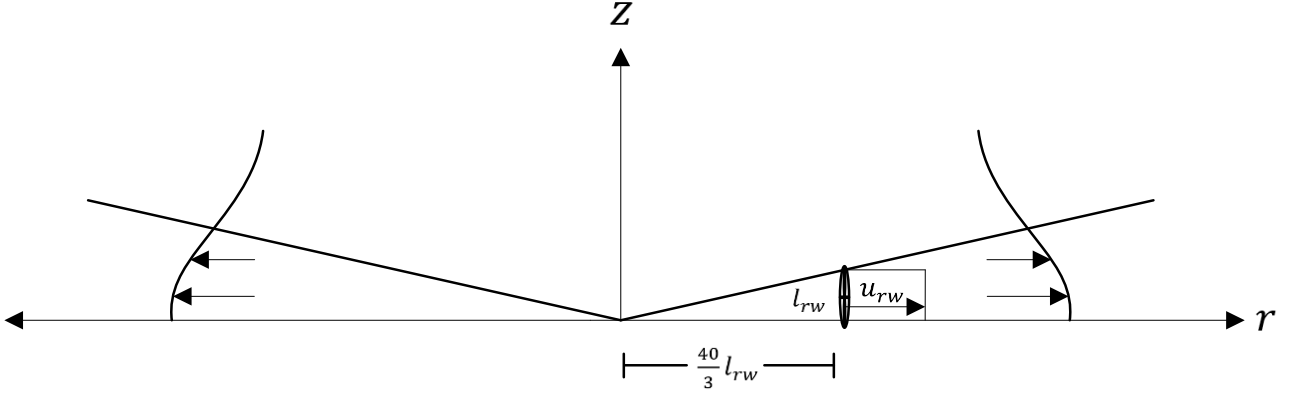

**Figure 3:** A radial wall jet emanating from a virtual point source. The real initial radial position of the jet is at $r = \frac{40}{3}l_{rw}$.

$$k_c = 0.1 u_c^2 \frac{(5l_c)^2}{z^2} e^{-70\xi^2} \tag{9}$$

where the 0.1 is the experimental relationship between the center turbulent kinetic energy and mean velocity squared (Rajaratnam 1976).

The other relevant jet form in this paper is the radial wall jet depicted in **Figure 3**, which occurs when a circular jet impinges on a flat wall. Through a similar analysis as for the circular jet, the mean velocity and turbulent kinetic energy of a radial wall jet may be expressed as

$$U_{rw} = u_{rw} \frac{\frac{40}{3}l_{rw}}{r} e^{-140\zeta^2} \tag{10}$$

$$k_{rw} = 0.14 u_{rw}^2 \frac{\left(\frac{40}{3}l_{rw}\right)^2}{r^2} e^{-140\zeta^2} \tag{11}$$

where $u_{rw}$ is the initial velocity of the jet, $l_{rw}$ is the initial size of the jet, and the self-preserving spread of the jet is $\zeta = z/r$, since the radial wall jet travels in the radial direction. The empirical power law scalings of Eqns. 10 and 11 ares an $r^{-1.1}$ and $r^{-2.2}$ dependence, respectively, because of energy losses due to wall friction (Launder 1983). However, in this work, we will only consider integer powers since the length scales of the spaces we will consider are not large enough for the decimal powers to produce significant deviations in Eqns. 10 and 11. The mathematical form of Eqns. 10 and 11 amounts to a radial wall jet with a virtual origin as shown in **Figure 3**. In experiment, this virtual source could be offset in the $\pm z$ direction due to the nature of the transformation between circular and radial wall jets while impinging on the wall. However, this offset is usually small, so Eqns. 10 and 11 are approximately valid.

Experiments in (Beltaos 1976) suggest that $\delta = 0.075r$ on average for any impingement angle, which gives rise to a different normalization condition and constant in the argument of the Gaussian than those for the circular jet. The 0.14 denotes that the magnitude of the turbulent kinetic energy with respect to the mean velocity squared is greater than that of the circular jet. This is likely to the wall friction (Launder 1983).

### 2.2 The Turbulent Diffusion Coefficient for Scalar Concentration

To simplify the evolution of the jets produced by fans in the space, the following two assumptions will be made:
1. The only regions of the jet which produce significant diffusion are the circular and first radial wall, so the complexities of subsequent forms may be ignored.
2. The jets minimally intersect in the region where their influence on diffusion is significant.

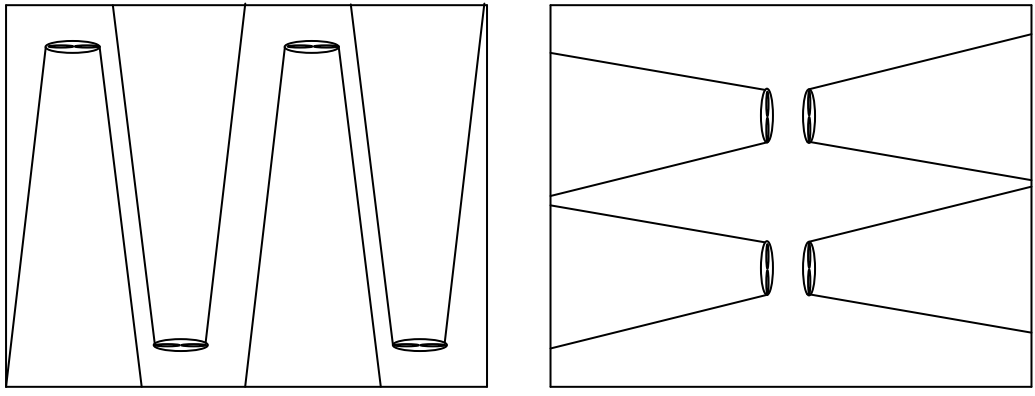

**Figure 4:** Examples of valid fan arrangements for assumption 2

The first assumption is valid since the magnitude of the turbulent kinetic energy decreases with distance, and so considering the complicated shapes of subsequent jets is unnecessary. The second assumption requires that the fans be spaced apart or facing different directions i.e. **Figure 4**.

The turbulent diffusion coefficient of a scalar concentration is defined as $D = c_v\sqrt{\langle k \rangle}l_D$ where $\langle k \rangle$ is the volume averaged turbulent kinetic energy, $l_D$ is the mixing length scale, and $c_v = 0.4$ is the von Karman constant. The simplest options for a mixing length scale are the length of the space and the length of the fan. (Foat 2020) demonstrated that the latter provides a significantly better correlation to experiment, and so in this paper the diffusion length scale will be set to the length of the fan ($l_D = l_c$). Note that by using the volume averaged turbulent kinetic energy of the jets versus the turbulent kinetic energy as a function of coordinates, the diffusivity is simplified to a constant. This implies that the turbulence produced by each fan is homogenous throughout the space. This is a good approximation, unless the volume of the space is much greater than the total volume of turbulence contained within the jets.

To simplify the calculation of $\langle k \rangle$, it will be demonstrated that the total turbulent kinetic energies in a cross-sectional area of the circular jet and radial wall jet it transforms into are constant and equal, implying that the total volume of turbulence they contribute to the space as a function of the distance they propagate is identical. This simplifies the analysis because we only need to use the initial conditions of the circular jet to determine $\langle k \rangle$, allowing for the calculation to be independent of the positions at which the fans are placed throughout the space. Again, energy losses due to wall friction will be neglected. A diagram of the transformation is depicted in **Figure 5**.

To show this, Eqns. 9 and 11 are integrated over the cross-sectional area of the circular and radial wall jets, respectively:

$$\int k_c dA = 0.11(u_c l_c)^2 \tag{12}$$

$$\int k_{rw} dA = 11.71(u_{rw} l_{rw})^2 \tag{13}$$

Eqns. 12 and 13 demonstrate that the turbulent kinetic energy in the cross-section of either jet is constant. To show that they are also equal, a relationship between $(u_c l_c)^2$ and $(u_{rw} l_{rw})^2$ must be found. Integrating the square of Eqns. 6 and 10 over the cross-sectional area of the circular and radial wall jets gives

$$\begin{aligned} M_c &= \int U_c^2 dA \\ &= 0.56(u_c l_c)^2 \end{aligned} \tag{14}$$

$$\begin{aligned} M_{rw} &= \int U_{rw}^2 dA \\ &= 59.16(u_{rw} l_{rw})^2 \end{aligned} \tag{15}$$

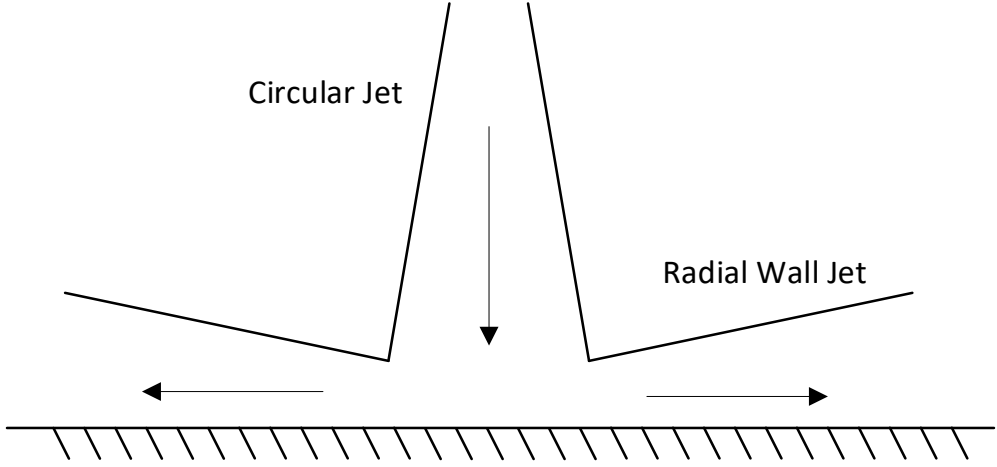

**Figure 5:** Transformation of a circular jet into a radial wall jet.

where $M_c$ and $M_{rw}$ are the momentum fluxes of the circular and radial wall jets, respectively. Since momentum flux is conserved, Eqns. 14 and 15 may be set equal, yielding $(u_{rw}l_{rw})^2 = 0.0095(u_c l_c)^2$. Substituting this into Eq. 13 gives

$$\int k_{rw} dA \approx 0.11(u_c l_c)^2 = \int k_c dA \qquad (16)$$

Therefore, since the initial conditions of the radial wall jet are the same as those for the circular jet when it impinges on the wall, the cross-sectional area of the circular jet and radial wall jet it transforms into are equivalent. Thus, $\langle k \rangle$ is independent of the positions of the fans in the room.

This also implies that $\langle k \rangle$ depends only on the initial conditions of the circular jet:

$$\langle k \rangle = \frac{1}{V_s}\int \left(\int k_c dA\right) dz = \frac{0.11(u_c l_c)^2}{V_s}\Delta z \qquad (17)$$

where $\Delta z$ is the distance the jet has propagated from the fan and $V_s$ is the volume of the space. The distance $\Delta z$ can be approximated by assigning a minimum turbulent kinetic energy to the jet, where "minimum" is relative to the strength of the diffusion it produces. Since the turbulent kinetic energy at the center of the jet initially is $k_c = 0.1 u_c^2$, with $u_c^2 \sim 1$ for most room fans, it is not unreasonable to suggest that the minimum turbulent kinetic energy is about an order smaller, $k_{min} \sim 0.01$. Using this, $\Delta z \approx 16\left[\frac{s}{m}\right]u_c l_c - 5 l_c$, where the bracketed term is the implicit units of the number 16. The diffusion coefficient for the $i^{th}$ fan in the room is thus:

$$D_i \approx 0.13 \frac{u_c^{3/2} l_c^{5/2}}{V_s^{1/2}} \left(16\left[\frac{s}{m}\right] - 5/u_c\right)^{1/2} \qquad (18)$$

For brevity, $u_c$ and $l_c$ will be renamed to $u$ and $l$, respectively. Eq. 18 is only accurate once the fans have established steady flow in the space, and thus have created approximately homogenous turbulence. Therefore, there will be a time delay, $t_{del}$, between the time the fan is switched on and the time steady flow is reached. This may be estimated by calculating the time it takes for the jet to propagate a distance $V_s^{1/3}$, which corresponds to the length scale of the space. If we have $N$ fans in the room that all have the same initial conditions, then we can imagine each fan effectively sub-diving $V_s$ into $N$ smaller parts, each with volume $V_s/N$. In this case, it would be more appropriate to calculate the time delay based on the length scale $(V_s/N)^{1/3}$. Using the equation for the mean velocity, $U_m = \frac{dz}{dt} = u\frac{5l}{z}$, the time delay is

$$t_{del} = \frac{(V_s/N)^{2/3}}{ul}\left(\frac{1}{10} + \frac{l}{(V_s/N)^{1/3}}\right) \qquad (19)$$

The time delay can also be a useful measure for determining if the initial conditions of the fans are suitable for the space volume. Ideal fan arrangements should have $t_{del} \sim 1s$. If $t_{del} \gg 1s$, then at least one of the following parameters should be increased: fan velocity $u$, size $l$, or the number of fans $N$. After the time delay has elapsed, Eq. 18 is valid and the diffusion coefficients of all the fans may be summed to form the total diffusion coefficient. Assuming all the fans have the same initial conditions, the total diffusion coefficient is just Eq. 18 multiplied by the number of fans:

$$D = 0.13 N \frac{u^{3/2} l^{5/2}}{V_s^{1/2}} \left(16 \left[\frac{s}{m}\right] - 5/u\right)^{1/2} \tag{20}$$

*2.3 The Mixing Time*

The model developed here involves the mixing of a single virus cloud produced by the cough of a high emitter. The virus cloud will be approximated as a sphere of radius $R$ at the center of the space, whose mixing is diffusive. The initial concentration of the virus droplets will be approximated as a constant inside the sphere with value $C_o$, and zero outside the sphere. Using the Green's function for the 3D diffusion equation, the peak concentration (at $r = 0$) is expressed as

$$C_p(t) = C(r = 0, t) = \frac{C_o}{(4\pi D t)^{3/2}} \int_0^{2\pi} d\theta \int_0^{\pi} d\varphi \int_0^R e^{-\frac{r^2}{4Dt}} r^2 \sin\varphi \, dr \tag{21}$$

The solution may be expanded for small $\frac{R}{\sqrt{Dt}}$, yielding:

$$C_p(t) \approx C_o \frac{R^3}{4\sqrt{\pi}(Dt)^{3/2}} \tag{22}$$

Assuming the fan arrangement is suitable for the space, the time delay defined by Eq. 19 is small compared to the mixing time and so it may be neglected. Thus, Eq. 22 may be inverted to yield the expression for the mixing time:

$$t_M = \frac{1}{D}\left(\frac{1}{\Gamma} \frac{3 V_{cl}}{16\pi^{3/2}}\right)^{2/3} \tag{23}$$

where $V_{cl} = \frac{4}{3}\pi R^3$ is the volume of the initial virus cloud and $\Gamma = \frac{C_p(t_M)}{C_o}$ is the goal concentration ratio, defined as the desired concentration at the mixing time divided by the initial concentration. Substituting Eq. 20 into Eq. 23 gives:

$$t_M \approx 0.8 \left(\frac{V_{cl}}{\Gamma}\right)^{2/3} \frac{V_s^{1/2}}{N u^{3/2} l^{5/2}} \left(16\left[\frac{s}{m}\right] - \frac{5}{u}\right)^{-1/2} \tag{24}$$

The amount of virus particles inhaled by a bystander depends on their time of exposure, $t_{ex}$, to the virus. Therefore, no matter how low the concentration of virus particles is in the space, there will always be a finite time that they will be safe from infection. Based on the parameters from various studies for the volume of a typical cough[1] (Lindsley 2012), the initial concentration of virus droplets in the cough of a high emitter[2] (Riediker 2020), the minimal dose for infection[3] (Ryan *et al* 2020), and the volume breathing rate[4] (Beardsell 2009), Eq. 24 becomes

$$t_M \approx 0.13 \left[\frac{m^2}{s^{2/3}}\right] \left(16\left[\frac{s}{m}\right] - \frac{5}{u}\right)^{-1/2} t_{ex}^{2/3} \frac{V_s^{1/2}}{u^{3/2} l^{5/2}} \tag{25}$$

---

[1] The average initial volume of virus clouds produced by coughs is $V_{cl} \approx 2.5 \times 10^{-3}$ m$^3$ according to Lindsley 2012.
[2] Riediker 2020 determined that high emitters produce coughs with virus particle concentration $C_o \approx 9.0 \times 10^7$ m$^{-3}$.
[3] Ryan *et al* 2020 found that 500 virus particles produced minimum viral RNA replication in ferrets.
[4] The resting breathing rate of a human is around $1.4 \times 10^{-4} \frac{m^3}{s}$ according to Beardsell 2009.

where $t_{ex}$ arises since the goal concentration ratio is $\Gamma \approx \frac{0.04}{t_{ex}}[s]$. For most values of $u$, the term $0.13 \left[\frac{m^2}{s^{2/3}}\right] \left(16 \left[\frac{s}{m}\right] - 5/u\right)^{-1/2} \approx 0.04 \left[\frac{m^{5/2}}{s^{7/6}}\right]$. The mixing time can be approximated as

$$t_M \approx \frac{t_{ex}^{2/3}}{25N} \sqrt{\frac{V_s}{u^3 l^5}} \qquad (26)$$

where the 25 has implicit units $\left[\frac{s^{7/6}}{m^{5/2}}\right]$. The units of this constant come from the substitution of numerical values for the cough volume, virus concentration, breathing rate, and $\Delta z$. For different initial conditions, this constant can change and so it should not be taken as a general result. The important takeaways from Eq. 26 are the order of magnitude of this constant $(\mathcal{O}(10))$ and the powers of the variables, which are general. Notice how the mixing time decreases more with respect to an increase in the fan size versus an increase in fan velocity. This emphasizes that one should focus on creating fan arrangements with larger fans rather than faster fans for a minimum mixing time. Moreover, the square root dependence on the space volume indicates that the mixing time does not increase strongly with the space volume.

## 3. Discussion of Applications of the Model

A useful application of Eq. 26 is to the case of a public restroom. Say a poorly ventilated, one-person restroom of volume 27 m³ is currently hosting a high emitter with COVID-19, while a bystander waits for the restroom's vacancy. The high emitter coughs once while using the restroom, and then leaves. If the bystander immediately enters the restroom, there is a significant probability that they would breathe in a critical dose of virus particles from the highly concentrated virus cloud. Now consider putting a fan with $u = 2$m/s, and $l = 0.5$m in the restroom (**Figure 6**). This is a suitable fan arrangement since the time delay is $t_{del} = 2.4$s, which is not $\gg 1$s. Since the time of exposure is roughly 5 minutes for a typical restroom use, the diffusion time is $t_M \approx 20$s. This means that if the fan is automatically switched on after the high emitter leaves, then the bystander needs to wait only 20 seconds before using the restroom. The mixing time is also lower than the ventilation time for a well-ventilated bathroom, which is around 3 minutes (Riediker 2020). Mixing reduces peak concentration on a much shorter time scale.

The problem becomes a bit more complicated if the fans are mixing the virus cloud while bystanders are in the space, for example, mixing in an office. First, the mixing time is nonzero, so there is a period when the air contains clouds with high concentrations of the virus. Second, more than one high emitter can cough, so modelling the mixing of just one cloud is inaccurate. Third, the high emitters may also cough periodically, which could increase the concentration of virus particles in the space to dangerous levels (if there is poor ventilation) even after mixing. Finally, the fans would create a constant loud noise that can be distracting.

The only solution to the first issue is to minimize the mixing time as much as possible. However, there will always be a period where occupants will be susceptible to inhaling large concentrations of the virus. The second issue can be addressed by assuming that high emitters likely are rare, and so it is unlikely that there will be multiple sources in proximity. The mixing of the virus clouds produced by multiple high emitters can thus be assumed to evolve independently in separate parts of the room, and Eq. 26 is still approximately valid. The only parameter that needs to change is the goal concentration for each cloud, which must be halved if there are two virus clouds being mixed at the same time. The third issue is problematic, because mixing cannot provide a solution. If the space has poor ventilation, no windows, etc. then the virus clouds will remain in the room for hours until their constituent droplets fall to the ground. The concentration of virus particles in the space, even after mixing, can continue to rise until it becomes dangerous for occupants to remain in the room for even short times. The only solution is to have a small enough ventilation time, so that it prevents the concentration of virus particles in the space (after mixing) from increasing to dangerous levels. The last problem is avoidable if the speed of the fans is kept minimal. This means that to increase mixing while maintaining low noise, one can only increase the size of the fan. This turns out to be not too much of a handicap since the mixing time (Eq. 26) decreases strongly as the fan size increases.

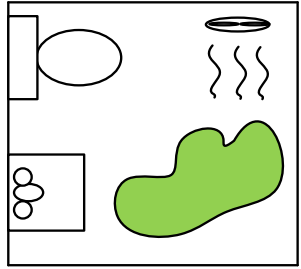

**Figure 6:** Mixing of a virus cloud in a one-person restroom

Consider an office space of volume 2000 m³ with 3 large fans that have $u = 3$ m/s, and $l = 1$ m. The time delay for this fan arrangement is about 8s, which is suitable for the space since it is not unreasonably larger than 1s. If $t_{ex} = 8$ hr (that of a typical workday), then $t_M \approx 3$min. Comparing this to the ventilation time of a typical office, which is around 20 min (Riediker 2020), we can conclude that this fan arrangement may improve safety for room occupants much more rapidly than the ventilation system alone. **Table 2** shows values of the mixing time (Eq. 26) for additional possible scenarios.

| Space Type | Typical Space Volume ($m^3$) | Typical $t_{ex}$ (hr) | $u$ (m/s) | $l$ (m) | N | $t_M$ (min) |
|---|---|---|---|---|---|---|
| **Classroom** | 300 | 1 | 2 | 0.5 | 1 | 6 |
| **Restaurant** | 1000 | 1 | 3 | 0.7 | 2 | 1 |
| **Auditorium** | 3000 | 2 | 3 | 0.5 | 5 | 2 |
| **Laboratory** | 1200 | 4 | 2 | 0.4 | 4 | 12 |
| **Plane Cabin** | 200 | 5 | 1 | 0.3 | 10 | 13 |

**Table 2:** Mixing time for different scenarios

## 4. Conclusion

There is ample evidence that the airborne transmission of COVID-19 is a significant mode of spreading in poorly ventilated, enclosed spaces. If present in these spaces, high emitters will produce virus clouds with high concentrations of virus particles which could instantly give passing bystanders a significant dose. A possible way to mitigate this problem is to mix the virus cloud in the space, increasing the cloud volume while simultaneously lowering the local virus particle concentration. This way, a bystander can remain in the room for a specified time of exposure before they are susceptible to infection. To demonstrate whether mixing is a viable option to reduce the spreading of COVID-19, an analytical model for the mixing of a virus cloud by fan arrangements which generate turbulent jets was developed, theoretically and by using the results of prior experiments. Two types of jets were considered: circular jets (emitted directly by the fans) and radial wall jets (created when a circular jet impinges on a wall). Approximating the turbulence as homogenous (after the time delay), the diffusivity was calculated for an arbitrary fan arrangement[5]. Using Fick's Law and data from coughs and SARS-CoV-2 experiments, the mixing time (Eq. 26) was calculated for a single virus cloud created by a high emitter's cough. Applications of Eq. 26 demonstrate that, for appropriate fan arrangements, the mixing time can be much smaller than poor or even typical ventilation times. It is crucial that mixing and ventilation be used together when the ventilation time is long. Mixing alone cannot prevent infection since virus concentrations can increase over time if there are high emitters in the space. This work suggests that mixing may contribute to improving the safety of poorly ventilated enclosed spaces under some conditions.

## Acknowledgements
This research was partially supported by U.S. DOE under Award No. DE-FG02-04ER54738.

## References
Anand S and Mayya Y 2020 Size distribution of virus laden droplets from expiratory ejecta of infected subjects *Sci Rep* **10** 21174

Beardsell I 2009 Get Through MCEM Part A: MCQs. *CRC Press*

---

[5] Standard Reynolds closure and similarity methods were used.

Beltaos, S 1976 Oblique Impingement of Circular Turbulent Jets *Journal of Hydraulic Research*

Bourouiba L, Dehandschoewercker E, and Bush J 2014 Violent expiratory events: On coughing and sneezing *Journal of Fluid Mechanics* **745**, 537-563

van den Brand J, Haagmans B, Leijten L, van Riel D, Martina B, Osterhaus A, and Kuiken T 2008 Pathology of experimental SARS coronavirus infection in cats and ferrets *Veterinary pathology* **45(4)** 551–562.

Duguid J 1946 The size and the duration of air-carriage of respiratory droplets and droplet-nuclei. *J. Hyg.* **44 (6)** 471–479

Foat T 2020 "A relationship for the diffusion coefficient in eddy diffusion based indoor dispersion modelling". *Building and Environment*

Gorbalenya A et al 2019 The species severe acute respiratory syndrome-related coronavirus: classifying 2019-nCoV and naming it SARS-CoV-2 *Nat. Microbiol.* **5** 536–544

Lai C 2019 Budgets of turbulent kinetic energy, Reynolds stresses, and dissipation in a turbulent round jet discharged into a stagnant ambient *Environmental Fluid Mechanics*

Launder B 1983 The Turbulent Wall Jet – Measurements and Modeling *Annual Review of Fluid Mechanics*

Lindsley W 2012 Quantity and Size Distribution of Cough-Generated Aerosol Particles Produced by Influenza Patients During and After Illness *Journal of Occupational and Environmental Hygiene*

Lu J, et al 2020 COVID-19 outbreak associated with air conditioning in restaurant, Guangzhou China *Emerg. Infect. Dis*.

Mingotti N, Wood R., Noakes C, and Woods A 2020 The mixing of airborne contaminants by the repeated passage of people along a corridor *Journal of Fluid Mechanics* **903** A52

Morawska L and Milton D 2020 It is time to address airborne transmission of COVID-19. *Clin. Infect. Dis.*

Nicas M, Nazarof W and Hubbard A 2005 Toward understanding the risk of secondary airborne infection: emission of respirable pathogens *J. Occup. Environ. Hyg*. **2(3)** 143–154

Parker-Pope T 2021 Can Vaccinated People Go to the Gym? *New York Times*

Pope S 2000 Turbulent Flows *Cambridge University Press*

Rajaratnam, N 1976 Turbulent Jets *Elsevier Scientific Publishing Company*

Riediker M 2020 Estimation of Viral Aerosol Emissions from Simulated Individuals with Asymptomatic to Moderate Coronavirus Disease 2019 *Jama Network Open*

Ryan K, et al 2020 Dose-dependent response to infection with SARS-CoV-2 in the ferret model: evidence of protection to re-challenge *Nature Communications*

Sonkin L 1951 The role of particle size in experimental airborne infection *Am. J. Hyg.* **53** 337–354

Stadnytskyi V, Bax C, Bax A and Anfnrud P 2020 The airborne lifetime of small speech droplets and their potential importance in SARS-CoV-2 transmission *Proc. Natl. Acad. Sci.*

Wells W 1934 On air-born infection. Study II. Droplet and droplet nuclei *Am. J. Hyg.* **20**, 611–618.

Wells W 1955 Airborne Contagion and Air Hygiene: An Ecological Study of Droplet Infection *Harvard University Press*